# A discrete, finite multiverse

## Alan McKenzie

*Lately of University Hospitals Bristol NHS Foundation Trust, Bristol UK*[*]

**Abstract**

The Many Worlds Interpretation (MWI) famously avoids the issue of wave function collapse. Different MWI trees representing the same quantum events can have different topologies, depending upon the observer. However, they are all isomorphic to the group of block universes containing all of the outcomes of all of the events, and so, in that sense, the group of block universes is a more fundamental representation. Different branches of the MWI tree, representing different universes in MWI, ultimately share the same quantum state in a common "ancestor" branch. This branching topology is incompatible with that of the Minkowski block universe; the resolution is to replace the branches with discrete, parallel block universes, each of which extends from the trunk to the outermost twigs. The number of universes in a branch is proportional to its thickness which, in turn, depends upon the absolute square of the probability amplitude for the state in that branch. Every quantum event may be represented by a "kernel" of universes, which is the smallest group of universes that will reproduce the quantum probabilities of the outcomes of that event. By considering the ratios of the probabilities of the outcomes of any event, it can be shown that the number of universes in every kernel must finite, as must be the total number of universes in the multiverse. Further, every universe in the multiverse must be finite both in space and time. Another consequence is that quantum probabilities must be rational, which suggests that quantum mechanics is only an approximation to a discrete theory. A corollary is that not every conceivable universe exists in the multiverse, no doubt to the disappointment of those who enjoy alternate-history novels.

1. **"Splitting" in the Many Worlds Interpretation**

In spite of the advice of his doctoral thesis advisor, John Archibald Wheeler, to remove the loaded references to "splitting" from the draft of his PhD thesis, Hugh Everett III managed to retain one mention of the word, in the context of the observer, in the finished work:

> "As soon as the observation is performed, the composite state is split into a superposition for which each element describes a different object-system state and an observer with (different) knowledge of it." [1]

---

[*] http://www.godel-universe.com/

Everett famously introduced the concept of the Many Worlds Interpretation (MWI) in order avoid the difficulty of wave-function collapse. The splitting envisaged by Everett is illustrated in Figure 1, which shows a few branches of the vast tree generated by the unitary evolution of the wave function of the whole universe.

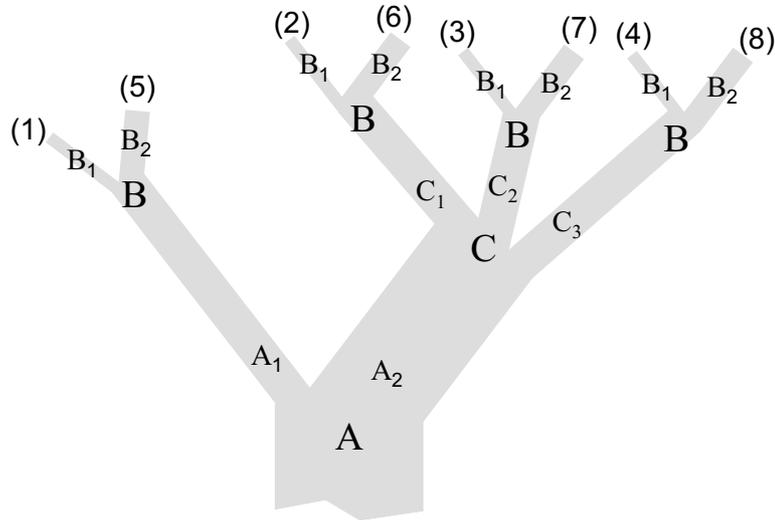

**Figure 1:** A MWI tree with three quantum events, A, B, and C. The outcomes of each of these events are indicated by the corresponding letter, numbered by a subscript.

Three quantum events are shown in the figure, labelled A, B and C. Sometimes in this paper we shall call these events "experiments" in the cause of tradition, and the environment in which the result of each experiment decoheres will be called the "Observer" (which may be just an interacting electron, not a sentient being!). The outcome of each experiment is labelled by a subscript. So, for instance, the outcome of experiment A may be either $A_1$ or $A_2$.

Possible sequences of events are implicit in the figure. For example, outcome $A_2$ is followed by outcome $C_1$, which, in turn, is followed by outcome $B_1$. Eight different sequences are represented in the tree, and these are shown schematically in Figure 2. Each of the eight possibilities is drawn within a Minkowski block universe. Alice, Bob and Cleo conduct the experiments A, B and C respectively.

The sequence $A_2$, $C_1$, $B_1$ mentioned above is contained in universe (2), and inspection of Figure 1 will confirm that the remaining seven sequences are also each represented in Figure 2. The numbers in brackets at the ends of the branches (the "twigs") in Figure 1 correspond to the universes listed in Figure 2. So, in a sense, the MWI tree in Figure 1 is *isomorphic* with the collection of block universes in Figure 2. Notice that experiment C is conditional on the outcome of experiment A being $A_2$ and not $A_1$, which is why Cleo does not appear in universes (1) or (5). (For the sake of generality, three branches, and not two, emerge from event C.)

A feature of MWI trees is that they may be drawn from the perspective of a particular Observer. The tree in Figure 1 captures Alice's experience. We suppose that Bob is sufficiently far from Alice, spatially, that she sees the result of the experiment performed by Cleo, who is spatially close to her, before Bob's light-cone can reach her.

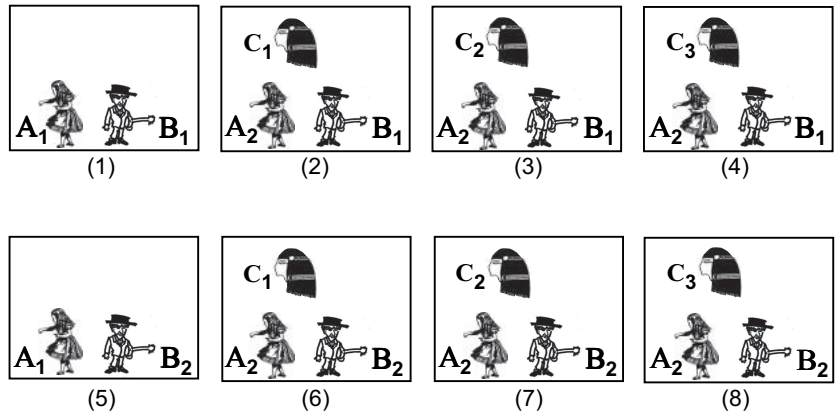

**Figure 2:** The eight possible outcomes of experiments A, B and C are shown in Minkowski block universes. Event C is conditional on the outcome of A being $A_2$.

Notice that Figure 1 features the same experiment B four times. If the tree had been drawn from Bob's perspective instead of Alice's, experiment B would only be seen once, but A and C would each appear twice – see Figure 3.

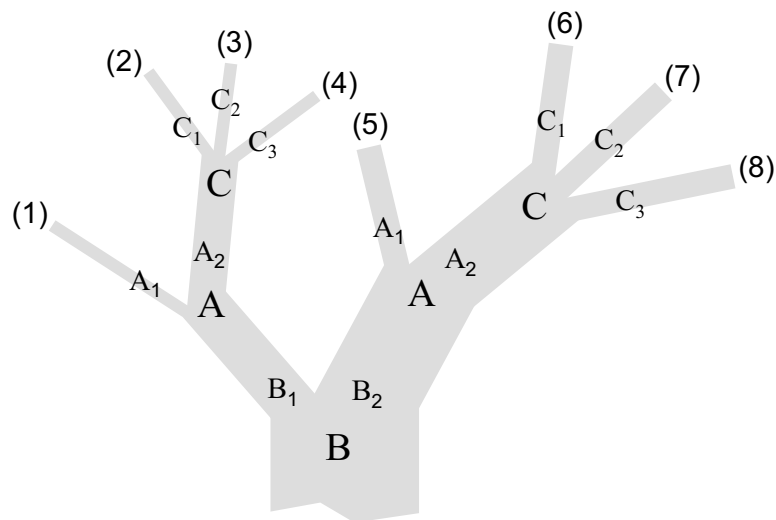

**Figure 3:** The same MWI tree as in Figure 1 but drawn from Bob's perspective.

Again, inspection of this figure will confirm that it contains all of the experimental sequences in Figure 1 – as seen in this case from Bob's perspective – and that, once more, it is isomorphic with respect to the eight possible sequences of outcomes of the three events in the collection of block universes in Figure 2. As before, the ends of the branches are labelled with the numbers in brackets that correspond to the eight possibilities in Figure 2.

Since the different perspectives of the MWI tree are isomorphic to the same collection of block universes, the latter may, in a sense, be regarded as a more fundamental representation. We shall return to this point in the next section.

Figure 4 shows the quantum states for a sample of the branches in Figure 1. The probability amplitude for each experimental outcome is indicated by the corresponding subscripted lower-case letter. So, for instance, the probability amplitudes for outcomes $A_1$ and $A_2$ of experiment A are $a_1$ and $a_2$ respectively. The state of the Observer is contained in its own subscript. Hence, for instance, $\left|\text{Obs}_{B_1 C_1 A_2}\right\rangle$ is an Observer (an environment) that has witnessed (decohered with) experimental outcomes $A_2$, $C_1$ and $B_1$ in the three experiments.

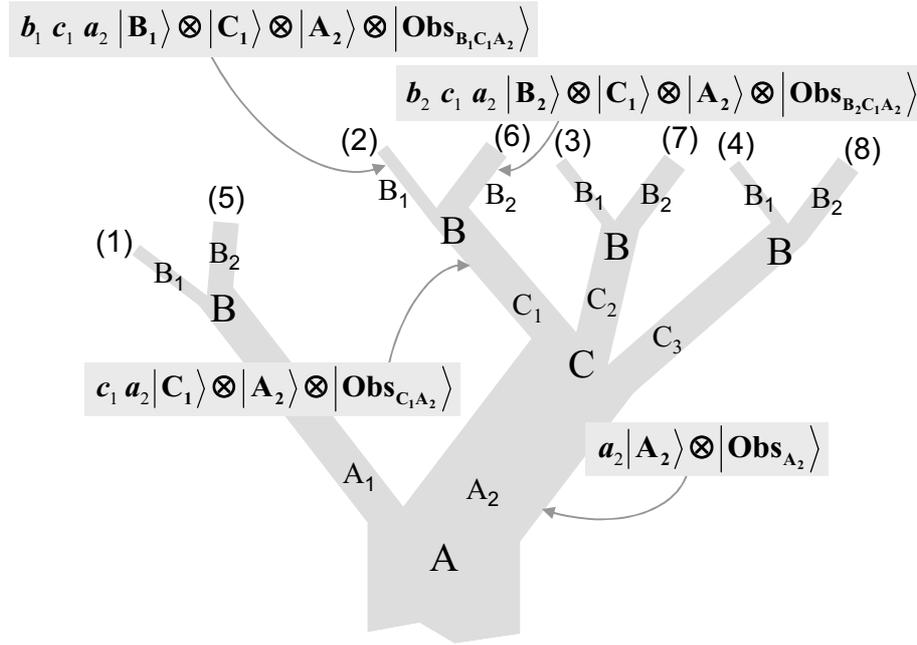

Figure 4: Quantum states for some branches. The components of each quantum state include states of outcomes of experiments in the ancestor branches multiplied by the corresponding probability amplitudes in lower-case letters. The subscripts of an environment – or Observer – state show the history that it has experienced.

The states of the twigs sprouting from the higher branches inevitably incorporate the history of all of their "ancestor" states. So, for instance, the two twigs labelled (2) and (6) that emerge from the B experiment each carry the imprint of the preceding C and A experiments, namely, $c_1 a_2 |C_1\rangle \otimes |A_2\rangle \otimes |Obs_{C_1A_2}\rangle$, while differing only in the two outcomes for the B experiment.

Regardless of Wheeler's nervousness over the "splitting" term, the provenance of each Observer's experience is clear from the diagram: each of the two different Observers, $|Obs_{B_1C_1A_2}\rangle$ and $|Obs_{B_2C_1A_2}\rangle$, in the two B branches shares the same ancestor, $|Obs_{C_1A_2}\rangle$, in the $C_1$ branch of the C experiment.

## 2. Reconciling the topologies of MWI and the block universe

The topology of this characteristic splitting is not allowed by the block-universe view of Minkowski space. Of course, the concept of a block universe is still challenged by some physicists (see discussion in [2]) but that is a debate for a different forum: the purpose of this paper is to work out the implications of a Many-Worlds picture that is compatible with block universes.

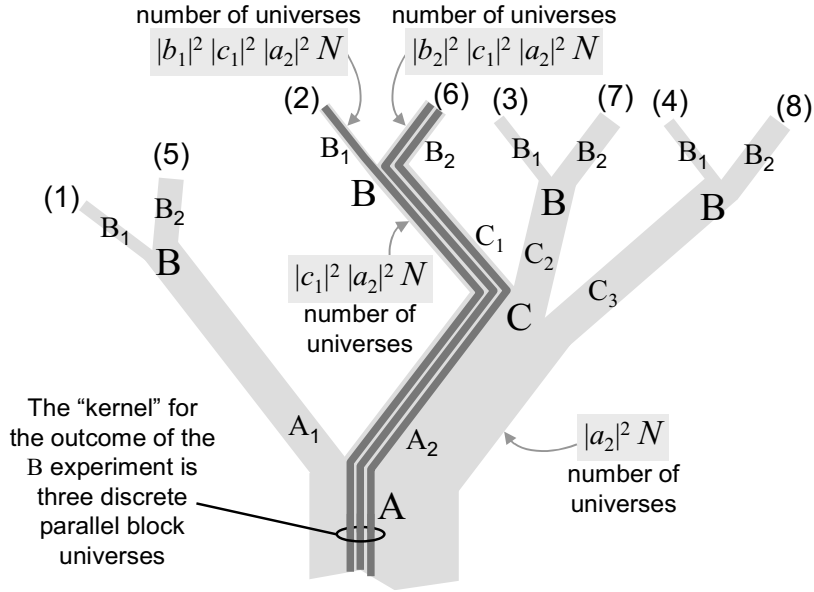

Figure 5: The MWI tree may be regarded as comprising many parallel filaments, each being a block universe. Three block universes are drawn as thick dark-grey lines.

Figure 5 illustrates how the topologies may be reconciled. Three indicative parallel block universes are shown, one being block universe (2) and two being block universe (6), as shown in Figure 2.

In the MWI the Observers in the (2) and (6) branches shared the *same* ancestor branch, $c_1 a_2 |C_1\rangle \otimes |A_2\rangle \otimes |Obs_{C_1A_2}\rangle$, containing the previous experiments, $C_1$ and $A_2$. However, in Figure 5 the three Observers' experiences of these previous experiments *are identical but separate*, because there is a copy of that branch, each

with its own copy of the Observer $|\text{Obs}_{C_1A_2}\rangle$, in each of the separate, discrete block universes. It is, perhaps, fitting that the resolution of the topology problem is solved by using a collection of parallel block universes which, as we noted in the previous section, is, in a sense, a more fundamental representation than the tree, since it is unchanged by an observer's perspective.

It will be useful to ask the question: how many parallel universes are there in the multiverse? In the MWI, the weighting, or "thickness", of each branch is proportional to the absolute square of the probability amplitude for the state in that branch [3]. In order for an Observer in a universe to find the same probability of an experimental outcome as that determined by the branch thickness, the number of universes in a branch must be proportional to the branch weighting, that is, proportional to the absolute square of the probability amplitude for the branch.

Suppose that the constant of proportionality is $N$, so that the number of universes in any branch is found by multiplying the absolute square of the probability amplitude for that branch by $N$. So the number of universes in each of the (2) and (6) branches is $|b_1|^2 |c_1|^2 |a_2|^2 N$ and $|b_2|^2 |c_1|^2 |a_2|^2 N$ respectively. Then, since $|b_1|^2 + |b_2|^2 = 1$, Observers in a fraction $|b_1|^2$ of all of the universes in these two branches will see an experimental outcome $B_1$, and, in a fraction $|b_2|^2$, Observers will see the outcome $B_2$. Of course, the quantum uncertainty experienced by the Observers arises because they do not know which universe they are in.

It is a feature of MWI trees that the total sum of the absolute squares of its probability amplitudes is conserved. To see this, note that the sum of the absolute squares of the probability amplitudes of the eight twigs (numbered (1) – (8) in Figure 5) is (taking the twigs in numerical order):

$|a_1|^2 |b_1|^2 + |a_1|^2 |b_2|^2 + |a_2|^2 |c_1|^2 |b_1|^2 + |a_2|^2 |c_1|^2 |b_2|^2 + |a_2|^2 |c_2|^2 |b_1|^2 + |a_2|^2 |c_2|^2 |b_2|^2$

$+ |a_2|^2 |c_3|^2 |b_1|^2 + |a_2|^2 |c_3|^2 |b_2|^2$

$= 1$ (because $|a_1|^2 + |a_2|^2 = |b_1|^2 + |b_2|^2 = |c_1|^2 + |c_2|^2 + |c_3|^2 = 1$)

So long as the number of universes in every branch is proportional to the absolute square of the probability amplitude for that branch, the fact that the total of the absolute squares is constant (unity) guarantees that the total number of universes in the multiverse remains constant despite the branching (as it should). Furthermore, since the absolute square of the probability amplitude of the beginning of the trunk of the tree must be unity (or it would not exist), the number of universes at that beginning must be $N$. So, in conclusion, the number of universes in the multiverse is the proportionality constant that is applied to every branch, namely $N$.

### 3. The number of universes in the multiverse

It is instructive to see whether there is a lower limit to $N$. We start by considering the number of universes needed to represent the outcomes of individual experiments. The ratio of the thicknesses of the branches from an individual experiment must be the same as the ratios of the absolute squares of the probability amplitudes for each of the possible outcomes of that experiment. So, for example, if outcome $B_2$ is twice as likely as that of $B_1$, then $|b_1|^2$ must equal ⅓ and $|b_2|^2$ must be ⅔. Therefore, the smallest number of universes that could represent this experiment alone, not considering any other experiment in the multiverse, is 3, with one universe allocated to the $B_1$ outcome and two to the $B_2$ outcome. Following the nomenclature of [4], let us call the smallest group of parallel universes that corresponds to a particular experiment, a *kernel*. So the kernel for the B experiment contains three universes: $K_B = 3$, with one in the $B_1$ branch ($K_{B1} = 1$), and two in the $B_2$ branch ($K_{B2} = 2$).

Taking another example, if the absolute squares of the probability amplitudes for the outcomes $C_1$, $C_2$ and $C_3$, namely, $|c_1|^2$, $|c_2|^2$ and $|c_3|^2$, are ⅓, ⅕ and 7⁄15 respectively, then the smallest group that reflects these probabilities has 5, 3 and 7 universes in these respective branches. The pattern becomes clear: the number of universes in each branch of a kernel is given by multiplying the absolute square of the probability amplitude for each of the experimental outcomes by the lowest common denominator (LCD) for the probabilities of each outcome. In this instance, the LCD is 15 and the numbers of universes are ⅓ × 15 = 5, ⅕ × 15 = 3, and 7⁄15 × 15 = 7 for the respective outcomes. So $K_C = 15$ with $K_{C1} = 5$, $K_{C2} = 3$ and $K_{C3} = 7$. We shall consider later the general case where the probability of an outcome cannot be expressed as a rational fraction.

In order to preserve the proportionalities of branch thicknesses for any given experiment, the number of universes in the trunk of the tree has to be the product of the kernels of every event in the tree. Then, proceeding upwards along the branches, the kernel for each experiment is divided into its components according to the relative thicknesses of the branches for that experiment. Figure 6 illustrates the process.

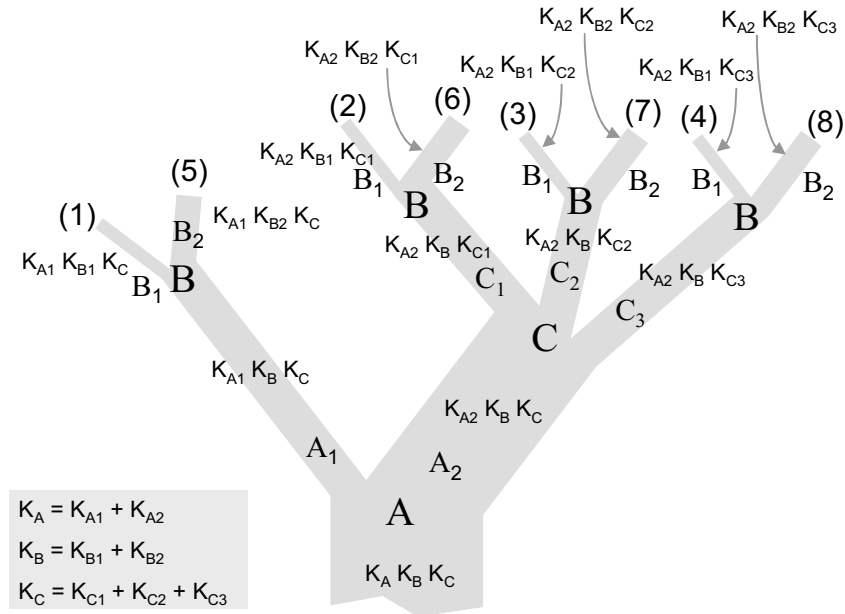

Figure 6: The smallest number of universes in any branch of the multiverse contains components of kernels from *every* branch in the multiverse.

From this, it is clear that the smallest possible number of universes in the multiverse, using the smallest possible number for each experiment (its kernel), is the product of the kernels of all of the experiments that take place in the multiverse. Notice that the repeated appearance of the same experiment on unconnected branches does not mean that its kernel has to be multiplied with each appearance (so that $K_B$ appears only once and not four times in the products).

In summary, then, the minimum number of universes in the multiverse, $N$, is given by the product of the kernels of all of the quantum events in the universe (which may be determined by inspection of diagrams like Figure 2). This means, of course, that the number of universes for every event has components of kernels of events in completely different branches of the multiverse. Hence, for instance, the twig containing outcome $B_1$ in universe (1) includes the kernel for event C in its product ($K_{A1} K_{B1} K_C$), although C occurs in a completely separate branch. Indeed, the number of universes containing *any* quantum event in the multiverse contains components of the kernels for *every* quantum event in the multiverse.

As we have seen, the branch kernels for the outcomes of a given experiment are in the same proportions to each other as those of the absolute squares of the relevant probability amplitudes (so, for instance, $K_{C1} : K_{C2} : K_{C3} \rightarrow |c_1|^2 : |c_2|^2 : |c_3|^2$). So none of the branch kernels can be infinite, because ratios of infinite quantities are not properly defined, whereas we can confidently verify the consistency of such ratios from the outcomes of multiply repeated C experiments (to within experimental error). This means that the complete kernel for any quantum event in the multiverse, being the sum of its finite branch kernels, cannot be infinite. Therefore $N$, which is the product of all of the kernels of all of the quantum events in the multiverse, cannot be infinite either. The number of universes in the multiverse is finite.

This conclusion, that the number of universes in the multiverse is finite, is unchanged even if that number is not $N$, but a multiple of $N$. It cannot be an infinite multiple, because that would mean, once again, expressing probabilities of infinite quantities as improperly defined ratios.

Of course, quantum mechanical probabilities calculated from the Schrödinger equation are generally irrational, and not expressible in the convenient ratios such as $\tfrac{1}{3}$, $\tfrac{1}{5}$ and $\tfrac{7}{15}$ that we used above. By definition, indeed, an irrational quantity cannot be expressed as the ratio of two finite numbers. However, the ratio of two infinite numbers is not properly defined. Therefore, Schrödinger-generated quantum probabilities cannot generally emerge from a multiverse of parallel block universes.

So, to summarise the argument so far:

1. The Many Worlds Interpretation of quantum mechanics circumvents the problem of wave-function collapse.

2. However, in turn, MWI introduces the difficulty that its topology is incompatible with that of the Minkowski block universe.

3. The resolution is to populate the MWI tree with discrete parallel block universes in proportion to the weightings of the branches.

4. However, that constrains calculated quantum probabilities to rational values only, and it also means that the number of parallel block universes in the multiverse is finite, albeit unimaginably vast.

## 4. A discrete theory

If the above arguments are accepted, then the difficulty with the irrational probabilities that emerge from the Schrödinger equation can only be resolved by regarding conventional quantum mechanics as an *approximation* to a discrete, digital theory (in a reversal of the normal situation where digital computers are used to simulate real and complex quantities). If the picture of quantum events in the MWI-block universe model is generalized to include quantum field theories extending, putatively, to one for the gravitational field, then this implies that a discrete theory should include general relativity.

If it could be shown that quantum physics or general relativity could not be replaced by such equivalent, discrete theories, then we should have to regard one or more steps in the above argument as invalid. It turns out, however, that physicists including Albert Einstein, Richard Feynman and Gerard 't Hooft among others have surmised that nature is fundamentally discrete [5].

Indeed, 't Hooft describes a quantum theory based upon integers [6], and Zahedi has reformulated the field equations of the strong, electromagnetic and gravitational forces (and other equations from quantum theory) in terms of integers and matrices: he prefaces his paper with a nice review of work in the field [5].

## 5. The multiverse and its component universes are finite

A corollary of the hypothesis that the number of parallel block universes in the multiverse is finite is that all of these universes must be finite in the time and spatial dimensions. If any universe were infinite in either time or space, then, from simple unitary evolution of the Schrödinger equation (or its discrete equivalent) within that universe, it would contain an infinite number of quantum events. These, in turn, would require an infinite number of kernels, which, from the previous discussion, would imply an infinite number of parallel block universes, which we have ruled out. A model of such a finite multiverse and its component block universes will be described in a subsequent paper.

Notice that a block universe is not generally the same thing as our observable universe, which is spatially finite by definition. While current data from, for instance, the Planck 2015 project [7] put the cosmological curvature parameter at zero to within half a percent (which implies a strong likelihood that the universe is infinite), some global topologies of the universe (such as Friedmann-Robertson-Walker models [8] allow the universe volume to be finite with a zero or even negative curvature [9]. So the above prediction that the universe is finite is not ruled out by current cosmological data.

Another corollary of a finite multiverse must be that, contrary to the popular device of alternate-history novels (e.g., Philip K Dick's *The Man in the High Castle*), the multiverse does not contain every conceivable universe.

If the premise for the Many Worlds Interpretation is accepted along with the concept of the block universe, then we may conclude:

1. The number of parallel block universes in the multiverse is finite.

2. Every universe in the multiverse is finite both in space and in time.

3. It is not true that every conceivable universe exists in the multiverse.

4. Quantum probabilities are never irrational.

5. Quantum theory is an approximation to an equivalent theory which must be discrete (this probably includes general relativity as well).

**6.      Testing the hypothesis**

A common response to the claim that MWI is not testable is that it is not a new theory but an interpretation of quantum mechanics, by which it stands or falls: every test of quantum mechanics is also a test of MWI. However, some of the above conclusions that result from combining MWI with parallel block universes may be tested, at least in principle (the block universe itself is testable – see [2]).

If, for instance, it turns out that the upper limit on the number of universes in a kernel is sufficiently low, then it might be possible to test the claim that quantum probabilities are always rational. Imagine an experiment, Z, which has two possible outcomes, $Z_1$ and $Z_2$, with conventionally calculated quantum-mechanical probabilities of ~1 and $10^{-25}$ respectively. The number of universes in the corresponding kernel components should therefore be $K_{Z1} = 10^{25}$ and $K_{Z2} = 1$. However, suppose that a kernel cannot exceed a limit of, say, $10^{20}$ universes, so that $K_{Z1} = 10^{20}$. In that case, $K_{Z2}$ might either be unity or zero, so that either the chance of detecting a $Z_2$ outcome in an experiment would be a factor of $10^5$ greater than predicted by conventional quantum mechanics or no such outcomes would be detected at all. Since experiments can be designed to detect such very low probabilities, a discrepancy between the measured and calculated probability would be evidence for the proposed hypothesis.

Another claim that might be tested is that every universe is finite both in space and in time. This claim runs counter to current cosmological thinking, as we said above, and so any new cosmological data to the contrary would support the hypothesis of a multiverse of discrete parallel block universes.

## References

[1]     Bryce Seligman DeWitt, R. Neill Graham, eds (1973). *The Many-Worlds Interpretation of Quantum Mechanics*, Princeton Series in Physics, Princeton University Press. ISBN 0-691-08131-X. Contains Everett's thesis: *The Theory of the Universal Wave Function*, pp 3–140.

[2]     McKenzie, A. (2016, March 28). Proposal for an experiment to demonstrate the block universe. arXiv:1603.08959 [physics.pop-ph].

[3]     Bryce Seligman DeWitt. (1972). The Many-Universes Interpretation of Quantum Mechanics, *Proceedings of the International School of Physics "Enrico Fermi" Course IL: Foundations of Quantum Mechanics*, Academic Press.

[4]     McKenzie, A. (2016, February 6). Some remarks on the mathematical structure of the multiverse. arXiv:1602.04247 [quant-ph].

[5]     Zahedi, R. (2016, May 5). On discrete physics: a perfect deterministic structure for reality – and a (direct) logical derivation of the laws governing the fundamental forces of Nature. arXiv:1501.01373v15 [physics.gen-ph].

[6]     't Hooft, G. (2014). Relating the quantum mechanics of discrete systems to standard canonical quantum mechanics. *Foundations of Physics*, 44(4), pp. 406-425.

[7]     Planck Collaboration: Ade, P. A. R., Aghanim, N., Arnaud, M. and others (2016, June 17). *Planck* 2015 results. XIII. Cosmological parameters arXiv:1502.01589v3 [astro-ph.CO].

[8]     Ellis, G. F. R and van Elst, H. (2008, September 2). Cosmological models (Cargèse lectures 1998) arXiv:gr-qc/9812046v5 .

[9]     Planck Collaboration: Ade, P. A. R., Aghanim, N, Armitage-Caplan, C. and others (2013, December 21). Planck 2013 results. XXVI. Background geometry and topology of the Universe. arXiv:1303.5086v2 [astro-ph.CO].